\documentstyle[aps,preprint,epsf]{revtex}

\newcommand{\vv}{{\rm v}}

\begin{document}

\title{High Order Perturbative QCD Approach to Multiplicity
Distributions of Quark and Gluon Jets}
\author{I.M. Dremin$^1$,  C.S. Lam$^2$, and V.A. Nechitailo$^1$}
\address{$^1$Lebedev Physical Institute, Moscow 117924, Russia \\
$^2$McGill University, Montreal H3A 2T8, Canada}

\maketitle

\begin{abstract} 
The second and third factorial moments of the multiplicity distributions of
gluon and quark jets are calculated up to the next-to-next-to-next-to leading
order in perturbative QCD, using the equations for generating functions.
The results are confronted with experimental data. A general discussion on
high order corrections revealed by such an approach is given.
Other possible corrections and approaches are discussed as well.
\end{abstract}

\section{Introduction}
The detailed properties of multiplicity distributions of gluon and quark jets
are studied in experiment nowadays \cite{pep,del,opa}. Theoretical
description of the data in the framework of perturbative QCD is rather
successful qualitatively, often at the accuracy  of 10--15$\%$
or better. Such an accuracy, obtained sometimes only in low-order
approximations, is somewhat surprising because the QCD expansion parameter is
rather large at present energies (about 0.5), so higher order
contributions should be estimated. It happens that different physical
quantities are sensitive in a different way to higher order corrections. A
thorough discussion of average multiplicities of gluon and quark jets is
given in [4-8] with respect to their dependence on these corrections. Here we
extend these studies to the widths and higher moments of multiplicity
distributions of gluon and quark jets, calculating them in the
next-to-next-to-next-to leading order approximation. Such an accuracy becomes
possible because the equations for the generating functions admit a
perturbative expansion. Besides finding out the appropriate coefficients, we
also discuss some general problems, revealed by such an analysis of average
multiplicities, widths, and higher moments.

\section{Widths of multiplicity distributions in perturbative QCD}
Any moment of the parton multiplicity distribution of gluon and quark jets in QCD can be obtained from the equations for the generating functions
\begin{eqnarray}
&G_{G}^{\prime }&= \int_{0}^{1}dxK_{G}^{G}(x)\gamma _{0}^{2}[G_{G}(y+\ln x)G_{G}
(y+\ln (1-x)) - G_{G}(y)] \nonumber \\
&+&n_{f}\int _{0}^{1}dxK_{G}^{F}(x)\gamma _{0}^{2}
[G_{F}(y+\ln x)G_{F}(y+\ln (1-x)) - G_{G}(y)] ,   \label{50}
\end{eqnarray}
\begin{equation}
G_{F}^{\prime } = \int _{0}^{1}dxK_{F}^{G}(x)\gamma _{0}^{2}[G_{G}(y+\ln x)
G_{F}(y+\ln (1-x)) - G_{F}(y)] ,                                   \label{51}
\end{equation}
where  $G_i$ are the generating functions of the multiplicity distributions
$P_n^{(i)}$ of gluon ($i=G$) and quark ($i=F$) jets, defined by
\begin{equation}
G_i(y,z)=\sum _{n=0}^{\infty }(z+1)^nP_n^{(i)}=\sum _{q=0}^{\infty }
\frac {z^q}{q!}\langle n_i\rangle ^qF_q^{(i)}.   \label{gen}
\end{equation}
In these expressions,  
$\langle n_i\rangle =\sum _{n=0}^{\infty }nP_n^{(i)}$ is the average
multiplicity, $z$ is an auxiliary variable, $y=\ln (p\Theta/Q_0)$ is the
evolution variable, $p, \Theta $ are the momentum and the opening angle of a
jet, $Q_0$=const ,
 $G^{\prime }(y)=dG/dy ,$ and $ n_f$ is the number of active flavors. Moreover,
\begin{equation}
\gamma _{0}^{2} (y)=\frac {2N_{c}\alpha _S(y)}{\pi } ,  \;\;
\alpha _S(y)=\frac {2\pi }{\beta_0y}\left ( 1-\frac {\beta _1\ln 2y}{\beta _0^2y}\right ),               \label{52}
\end{equation}
\begin{equation}
\beta _0=\frac {11N_c-2n_f}{3}, \;\;  \beta _1=\frac {17N_c^2-n_f(5N_c+3C_F)}{3},
\end{equation}
$\alpha _S$ is the running coupling strength, $N_c$ is the number of colours, 
and $C_{F}=  (N_{c}^{2}-1)/2N_{c}=4/3$ in QCD. The argument of $\gamma _0^2$ 
in the integrals is chosen to be $y+\ln x(1-x)$, as determined
 by the transverse momentum of partons at the splitting vertex. The kernels of the equations are
\begin{equation}
K_{G}^{G}(x) = \frac {1}{x} - (1-x)[2-x(1-x)] ,    \label{53}
\end{equation}
\begin{equation}
K_{G}^{F}(x) = \frac {1}{4N_c}[x^{2}+(1-x)^{2}] ,  \label{54}
\end{equation}
\begin{equation}
K_{F}^{G}(x) = \frac {C_F}{N_c}\left[ \frac {1}{x}-1+\frac {x}{2}\right] .
\label{55}
\end{equation}
The normalized factorial moment of any rank $q$ can be obtained by
differentiation
\begin{equation}
F_q^{(i)}=\frac {1}{\langle n_i\rangle ^q}\frac {d^qG_i}{dz^q}\vert _{z=0}, \label{fq}
\end{equation}
or, equivalently, by using the series (\ref {gen}) and collecting the terms
with equal powers of $z$ on both sides of the equations (\ref{50}), (\ref{51}).

The normalized second factorial moment $F_2$ defines the width of the
multiplicity distribution, and is related to its dispersion
$D^{2}=\langle n^2\rangle -\langle n\rangle ^2$ by the formula
\begin{equation}
D^{2}=(F_{2}-1)\langle n\rangle ^{2}+\langle n\rangle =K_{2}\langle n\rangle ^{2}
+\langle n\rangle ,     \label{disp}
\end{equation}
where $K_2$ is the second cumulant.

The second factorial moments normalized to their own average
multiplicities squared are
\begin{equation}
F_2^G =\frac {\langle n_G(n_G-1)\rangle }{\langle n_G\rangle ^2}, \;\;\;\;
F_2^F =\frac {\langle n_F(n_F-1)\rangle }{\langle n_F\rangle ^2}. \label{fgff}
\end{equation}

Let us write down their perturbative expansions up to $\gamma _0^3$-terms as
\begin{equation}
F^G_2=\frac{4}{3}(1-f_1\gamma_0-f_2\gamma_0^2-f_3\gamma _0^3),  \label{f2g}
\end{equation}
\begin{equation}
F^F_2=(1+\frac{r_0}{3})(1-\phi_1\gamma_0-\phi_2\gamma_0^2-\phi _3\gamma _0^3), \label{f2f}
\end{equation}
where $r_0=N_c/C_F$   determines the asymptotical value of the ratio of
multiplicities in gluon and quark jets (see (\ref{ra}) below). It is equal
to 9/4 in QCD  and to 1 in SUSY QCD. Actually, the asymptotical values of
$F_2^G$ and $F_2^F$ in front of the brackets in (\ref{f2g}), (\ref{f2f}) are
found out from the equations below by equating the leading terms on both
sides. However, we have inserted  their explicit expressions directly here to
simplify further notations.

Using the Taylor series expansion of $G$'s as proposed in \cite{dr}, one can
rewrite the Eqs. (\ref{50}), (\ref{51}) as
\begin{eqnarray}
\frac {1}{\gamma _0^2}[\ln G_G]^{''}=G_G-1-2G_G^{'}\vv_1+G_G^{''}\vv_2+
0.5G_G^{'''}\vv_3+
\nonumber
\\
\left (\frac {G_G^{'2}}{G_G}\right )^{'}\vv_{12}+4B\gamma _0^2[(G_G-1)\vv_1-
G_G^{'}\vv_2]+
\nonumber
\\
\frac {n_f}{4N_c} [ \left (\frac {G_F^2}{G_G}\right )^{'}\vv_4+
2\left (\frac {G_FG_F^{'}}{G_G}\right )^{'}\vv_5
+\left (\frac {G_FG_F^{''}}{G_G}\right )^{'}\vv_6+\left (\frac
{G_F^{\prime 2}}{G_G}
\right  )^{'}\vv_{13}-
\nonumber
\\
2B\gamma _0^2\left  ( \left (\frac {G_F^2}{G_G}-1\right )\vv_4+2\left (
\frac {G_FG_F^{'}}{G_G}+
\left ( \frac {G_F^2}{G_G}\right )^{'}\right )\vv_5\right ) ] ,    \label{gg''}
\\
\frac {r_0}{\gamma _0^2}[\ln G_F]^{''}=G_G-1-G_G^{'}\vv_7-G_G^{''}\vv_8-
0.5G_G^{'''}\vv_9-
\nonumber
\\
\left (\frac {G_GG_F^{'}}{G_F}\right )^{'}\vv_{10}-0.5\left
(\frac {G_GG_F^{''}}{G_F}\right )^{'}\vv_{11}-
\left (\frac {G_G^{'}G_F^{'}}{G_F}\right )^{'}\vv_{14}+
\nonumber
\\
2B\gamma _0^2[(G_G-1)\vv_7+2G_G^{'}\vv_8+
(G_G^{'}+\frac {G_GG_F^{'}}{G_F})\vv_{10}].      \label{gf''}
\end{eqnarray}
The terms up to the third derivative of $G$ are kept everywhere because
each derivative gives rise to the factors $\gamma $ or $\gamma _F$ as seen 
from their definition below in 
Eq. (\ref{an}), and we are interested in corrections up to $\gamma _0^3$.
We use $B=\beta _0/8N_c, \; B_1=\beta _1/4\beta _0N_c$.
The corresponding expansions for the anomalous QCD dimensions $\gamma $ and
$\gamma _F$ defined as
\begin{equation}
\langle n_G\rangle \propto \exp (\int ^y\gamma (y')dy'), \;\;
\langle n_F\rangle \propto \exp (\int ^y\gamma _F(y')dy' )     \label{an}
\end{equation}
are
\begin{equation}
\gamma=\gamma _0(1-a_1\gamma _0-a_2\gamma _0^2-a_3\gamma _0^3)   \label{gam}
\end{equation}
with $\gamma _F=\gamma -r'/r$, where
\begin{equation}
r=\langle n_G\rangle /\langle n_F\rangle =r_0(1-r_1\gamma _0-r_2
\gamma _0^2-r_3\gamma _0^3).    \label{ra}
\end{equation}
All the coefficients $a_i, r_i$ have been calculated and tabulated in
\cite{cd}. The integrals $\vv_i$ and terms in the right-hand sides of
equations (\ref{gg''}), (\ref{gf''}) proportional to
$\langle n_G\rangle ^2z^2/2$ are given in the Appendix. The corresponding
terms in the left-hand sides can be written as
\begin{equation}
[\ln G_G]''=\frac {\gamma _0^2\langle n_G\rangle ^2z^2}{2}\frac {4}{3}[1+
\sum _{1}^{3}M_n\gamma _0^n];  \;\; r_0[\ln G_F]''=\frac {\gamma _0^2\langle
n_G\rangle ^2z^2}{2}
\frac{4}{3}[1+\sum _{1}^{3}N_n\gamma _0^n],    \label{mn}
\end{equation}
where
\begin{eqnarray}
M_1=-(2a_1+0.5B+4f_1),
\nonumber
\\
M_2=a_1^2-2a_2+Ba_1-4f_2+4f_1(2a_1+1.5B),
\nonumber
\\
M_3=2a_1a_2-2a_3+0.5B(3a_2-B_1)-4f_3+2f_2(4a_1+5B)
\nonumber
\\
-4f_1(a_1^2-2a_2+2Ba_1+0.75B^2),
\nonumber
\\
N_1=-2a_1-0.5B-(1+\frac{3}{r_0})\phi _1+2r_1,
\nonumber
\\
N_2=a_1^2-2a_2+Ba_1-(1+\frac {3}{r_0})\phi_2 +(1+\frac {3}{r_0})\phi _1
(2a_1+1.5B-2r_1)+
\nonumber
\\
2r_2+r_1(3r_1-4a_1-3B),
\nonumber
\\
N_3=2a_1a_2-2a_3+0.5B(3a_2-B_1)-(1+\frac {3}{r_0})\phi _3 +
(1+\frac {3}{r_0})\phi _2(2a_1+
\nonumber
\\
2.5B-2r_1)-(1+\frac {3}{r_0})\phi _1[a_1^2-2a_2+2Ba_1+0.75B^2+
\nonumber
\\
2r_2+r_1(3r_1-4a_1-5B)]
+2r_3+r_2(6r_1-4a_1-5B)+
\nonumber
\\
r_1(4r_1^2-7.5Br_1-6a_1r_1+4Ba_1+2a_1^2-4a_2+1.5B^2).  \label{mnc}
\end{eqnarray}
The terms with the same power of $\gamma _0$ in expressions on both sides
should be equal. Therefore, one gets
\begin{eqnarray}
f_1=\frac{1}{3}[4\vv_1-2a_1-B/2
-\frac{n_f}{4N_c}\vv_4\left(1-\frac{5}{r_0}+\frac{6}{r_0^2}\right) ],
\\
\phi_1=\frac{1}{1+3/r_0}(f_1+2r_1+2\vv_7-2a_1-B/2),
\\
f_2 = \frac{2}{3}f_1(4a_1-2\vv_1+3B) +
     \frac{1}{3}( a_1^2 - 2a_2 - 4\vv_2 - 4a_1 \vv_1 + B a_1 - 4 B \vv_1 )
\nonumber
\\
 +\frac{n_f}{4N_c}\left\{
 \vv_4 [\phi_1\frac{1}{r_0}(\frac{1}{3}+\frac{1}{r_0}) - \frac{2}{3} f_1
 + 4\frac{r_1}{r_0}(\frac{1}{3}-\frac{1}{r_0})
                + (a_1+B)(\frac{1}{3}-\frac{5}{3r_0} + \frac{2}{r_0^2}) ]
\right.
\nonumber
\\
\left.
- 4\frac{\vv_5}{r_0^2}(1-\frac{r_0}{3})
\right\},
\\
\phi_2 = \phi_1( 2a_1 + \frac{3}{2}B-2r_1 ) +
     \frac{1}{1+3/r_0}(
     f_2 + 2 r_2 + 3 r_1^2 - 2 a_2 - 4 a_1 r_1
\nonumber
\\
+ a_1^2 - 3 B r_1 + B a_1 - 2\vv_7 (f_1+a_1+B) + 4 \vv_8 + \frac{4}{r_0}
\vv_{10}) \ \end{eqnarray}

\begin{equation}
f_3=\frac {1}{3}[d_f+\sum _{i=1}^{6}d_i\vv_i+d_{12}\vv_{12}+d_{13}\vv_{13}],
  \label{f3}
\end{equation}
where $d_f=M_3+4f_3$,
\begin{equation}
\phi _3=\left (1+\frac {3}{r_0}\right )^{-1}[f_3+d_{\phi }+\sum _{i=7}^{11}d_i
\vv_i+d_{14}\vv_{14}],  \label{p3}
\end{equation}
where $d_{\phi }=N_3+(1+\frac {3}{r_0})\phi _3$.  The coefficients $d_i$ are
given in the Appendix.
The numerical values  of $f_i, \, \phi_i$ for different number of active
flavors are shown in the Table 1.

\bigskip

Table 1

\bigskip

\begin{tabular}{|c|c|c|c|c|c|c|}
\hline
$n_f$ & $f_1$ & $f_2$&$f_3$ & $\phi_1$ & $\phi_2$&$\phi _3$\\
\hline
3 & 0.364 & -0.0279 &  0.795 & 0.637 & -0.276 &  2.12      \\
\hline
4 & 0.358 & -0.0457 & 0.740 & 0.631 & -0.286 &  2.04  \\
\hline
5 & 0.352 & -0.0629 & 0.689 & 0.625 & -0.295 &  1.95        \\
\hline
S & 0.313 & 0.310 &  -0.120    & 0.313 & 0.310 & -0.120  \\
\hline
\end{tabular}

\bigskip

Herefrom it is easy to see that the asymptotical $(\gamma _{0}\rightarrow 0)$
values of $F_{2}^{G}$ and $F_{2}^{F}$ are different as has been known since
long ago \cite{and}:
\begin{equation}
F_{2, as}^{G}=\frac {4}{3}, \;\;\;\; F_{2, as}^{F}=1+\frac {r_0}{3}=
\frac {7}{4}.  \label{fas}
\end{equation}
The experimental values \cite{opa} for 41.8 GeV gluon jets, $F_2^G=1.023\pm
0.008\pm 0.011$,
and for 45.6 GeV uds quark jets, $F_2^F=1.0820\pm 0.0006\pm 0.0046$, are much
lower than the above asymptotical limits. If one accepts, however, the
effective value of $\alpha _S$, averaged over all the energies of partons
during the jet evolution, to be 0.2, then one gets
$F_2^G(NLO)\approx 1.039$ and $F_2^F(NLO)\approx 1.068$, 
by taking into account only
the first correction proportional to $\gamma _0$. This is quite close to the
experimental results \cite{opa}. For $\alpha _S=0.12$, one gets about 10$\%$
higher values of $F_2$'s. Thus we conclude that the NLO-approximation
describes the widths at $Z^0$-energies within 10--15$\%$ accuracy.

At lower energies the widths should be slightly smaller due to the slow
increase of $\alpha _S$ and somewhat smaller effective values of $n_f$
leading to larger $f_1, \phi _1$. Using the values of $\alpha _S$ given
in the PDG-data \cite{pdg} and $n_f=4$, we predict the energy dependence of
NLO-values of second factorial moments shown by the solid
curves in Fig. 1.

\noindent
The additional dots at the ends of these curves demonstrate the effect due
to possible change of effective values of $n_f$ to 3 at $Q=10$ GeV and to
5 at $Q=90$ Gev, where $Q$ is the total energy as in \cite{pdg}. They
show how small is the indefiniteness imposed by effective number of
active flavours which is the only free parameter in such an approach .
The curves indicate that the widths are closer to
Poissonian ones at lower energies. Qualitatively, it agrees with experimental
trends observed by the DELPHI collaboration \cite{del}. The coupling strength
$\alpha _S$ changes within the interval of $Q$ shown here from 0.18 to 0.12.
The choice of the energy scale for jets is, however, highly nontrivial
(see, e.g., \cite{del,opa}). Therefore we do not plot experimental values 
here, just claiming the qualitative agreement within 15$\%$ accuracy.

The cut-off of the integration region at $\varepsilon=e^{-y}\approx e^{-2\pi /
\beta _0\alpha _S}$ from below and at $1-\varepsilon $ from above is not very
crucial at present energies as seen from the dashed curves in Fig. 1. It
diminishes the correction terms and, therefore, slightly increases the
widths and flattens their energy dependence. Thus, the role of the power
corrections is not very important.

Unfortunately, the higher order corrections do not improve our estimates. On
the contrary, the 2NLO-term is positive and tends to violate slightly the
agreement with experiment while 3NLO corrections are negative and so large
that lead even to sub-Poissonian widths of distributions ($F_2<1$) for
$\alpha _S=0.2$. These terms are approximately equal to NLO corrections due
to large values of $f_3$ and $\phi _3$. The inception of such large values
can be traced to rather large contributions of integrals containing $\ln ^2x$,
i.e. to the region of very soft gluons. Thus the cut-off at $e^{-y}$ and
$1-e^{-y}$ becomes more important for these terms.

The 3NLO-corrections are overestimated due to the adopted Taylor series 
expansion with the assumption $y\gg \vert \ln x\vert $ which is invalid for
soft gluons. For example, the $k_t$-dependence of the coupling strength is
transformed so that
\begin{equation}
\alpha _S\propto \frac {1}{y+\ln x(1-x)}\approx \frac {1}{y}\left (1-
\frac {\ln x(1-x)}{y}\right ),  \label{yx}
\end{equation}
and the second term becomes infinitely large at $x\rightarrow 0$. The cut-off
at $x=e^{-y}$ leads to a factor of 2 only. Thus the above expansion implies
some special presumption about the coupling strength behavior in the
nonperturbative region as well as its modification at the limits of the
perturbative one.
 The series (\ref{f2g}), (\ref{f2f}) are with the sign-changing and increasing
(in modulus) terms. From the values of $f_i, \phi _i$ in Table 1, 
one concludes that higher order
terms are more important for the width of a quark jet compared with 
a gluon jet.

The slopes of the widths are especially sensitive to these higher order terms
because each of them is enlarged by the factor  $n$ when differentiating
$\gamma _0^n$. Thus 3NLO contribution is about 3 times larger than the NLO
term in the slopes of widths. It demonstrates that any precise quantitative
estimates of slopes become impossible. In particular, at present energies one
cannot trust NLO estimates of these slopes as being small: $F_2^{G'}(NLO)
\approx 0.04; \; F_2^{F'}(NLO)\approx 0.092$ at $\alpha _S=0.2$. However, one
can predict the asymptotical value of the ratio of slopes as
\begin{equation}
\frac {(F_{2}^{G})^{'}_{as}}{(F_{2}^{F})^{'}_{as}}=\frac {16f_1}{21\phi _1}
\approx 0.43,   \label{rfpr}
\end{equation}
which surely coincides with their NLO ratio. It demonstrates that the second
factorial moment of quark jets approaches its asymptotical value faster than
for gluon jets.

Let us stress that all slopes and curvatures in pQCD are related to the running
property  of the QCD coupling constant since they are proportional to its
derivatives which are equal to zero for a fixed coupling constant. It is
interesting to note that the two-loop term in $\alpha _S$ proportional to
$\beta _1$ contributes only to the left-hand side of Eqs (\ref{mn}), namely,
to the coefficients $M_3$ and $N_3$ in (\ref{mnc}), and its role is very mild
there (about 1--2 $\%$). Thus it can be accounted with high accuracy
considering it only in the expressions for the expansion parameter $\gamma _0$.

One can also easily check that all the relations of SUSY QCD (where $n_f =N_c =
C_F$) are valid for all the coefficients shown above (e.g.,
$F_{2}^{G}=F_{2}^{F}$ etc.).
The SUSY values of $f_i=\phi _i$ are  shown in the lower line of the Table 1
marked by S. The asymptotical SUSY values of $F_2$ are equal to 4/3.

\section{Third moments of the multiplicity distributions}

The system of equations (\ref{gg''}), (\ref{gf''}) valid up to 3NLO-terms
of $\gamma _0^3$ was also applied by us to calculation of third moments of
the multiplicity distributions. It was done by equating the terms proportional
to $z^3$ on both sides of the equations and writing down the third factorial
moments (defined by Eq. (\ref{fq}) at $q=3$) as
\begin{equation}
F_3^G=h_0(1-\sum _{i=1}^{3}h_i\gamma _0^i); \;\;
F_3^F=g_0(1-\sum _{i=1}^{3}g_i\gamma _0^i).  \label{f32}
\end{equation}
Proceeding in the same way as done above, one gets $h_0=9/4, \; g_0=1+r_0+
r_0^2/4$ and the values of the coefficients $h_i, g_i$ in (\ref{f32}) shown
in the Table 2. The asymptotic limit of the third moment of quark jets is
about twice larger than that of gluon jets.
The analytic expressions are too lengthy to be presented here.

\bigskip

Table 2

\bigskip

\begin{tabular}{|c|c|c|c|c|c|c|}
\hline
$n_f$ & $h_1$ & $h_2$ & $h_3$ & $g_1$ & $g_2$ & $g_3$ \\
\hline
3     & 0.986  & -0.342 & 2.49 & 1.61 & -1.58 & 7.74   \\
\hline
4     & 0.972 & -0.380 & 2.36 & 1.60 & -1.59 & 7.54 \\
\hline
5     & 0.957 & -0.417 & 2.25 & 1.59 & -1.60 & 7.34 \\
\hline
S     & 0.844 & 0.722 & -1.09 & 0.844 & 0.722 & -1.09 \\
\hline
\end{tabular}

\bigskip

Comparing $f_i, \phi _i$ with $h_i, g_i$ one concludes that the corrections
increase for higher moments even in the NLO-approximation. Moreover, at present
values of $\gamma _0\approx 0.5$ they are rather large.
Let us note the similarity in the structure of corrections for widths and third
moments. They alternate in sign, and third coefficients are larger than the
first ones. It is an indication on the sign-alternating asymptotic series,
and Borel summation can be effective here. The increase of the coefficients
originates from the terms containing the integrals of the type $\int _0^1
\ln ^nxdx \propto n!$. The termination of the cascade at $\varepsilon =e^{-y}$
leading to power corrections becomes more important below $Z^0$. This reminds of
the situation with renormalons (see, e.g., \cite{zak}).

In SUSY QCD the asymptotical values of third moments are equal to 9/4. The
first correction given by $h_1$(SUSY)=$g_1$(SUSY)=0.844 is almost as large as
for ordinary gluon jets. It is similar to the correction for second moments.
However, NNLO and 3NLO-terms for moments in SUSY QCD differ drastically from
those for ordinary jets both in absolute values and signs as seen from the
Tables 1 and 2. It demonstrates their sensitivity to the value of $r_0$
which is drastically different in the two cases.

The similar procedure can be used for higher rank moments as well.

\section{General discussion and conclusions}
The equations (1), (2) are dealing with probabilities and, therefore, are of
classical nature. However, the quantum-mechanical interference has been
accounted in the angular ordering effect. Nevertheless, there is no proof of
their validity at all orders. The approach advocated above treats
these equations as the kinetic equations in QCD for partonic
cascade processes. Implicitly we have assumed that these equations describe
the cascade down to extremely low energies of partons by imposing the limits
0 and 1 of integration in the shares of energy. Thus the non-perturbative
region of soft partons was assumed to be described in the same manner as the
perturbative cascade. Probably, one should include into the consideration
only the perturbative region by the requirement that the evolution parameters
under the integral in (1), (2) stay always positive i.e. by introducing the
cut-off at $\varepsilon $
and $1-\varepsilon $ where $\varepsilon =e^{-y} $. Such a modification would
lead to the power-like corrections which can be neglected asymptotically but
contribute at present energies. Their role was considered for average
multiplicities in \cite{cd}. They are not very important for the moments in
the NLO-approximation as shown above.
However, more thorough treatment is needed,
especially, in view of rather large contribution of soft gluons to $f_3$ and
$\phi _3$.

Other  analytic solutions of these equations  different from the perturbative
one as well as nonperturbative modifications of the equations 
\cite{imd,los,ekg} can be looked for. 
Especially interesting would be to learn more about
the singularities of the generating functions in the $z$-plane which are
known up to now for the leading order solution only.

Leaving this program for future studies, we can now compare the results
obtained from perturbative solutions of the equations for average
multiplicities  [3-7] and for widths of the multiplicity distributions of
gluon and quark jets up to 3NLO approximation of pQCD. Leading order
predictions for any quantity are quite far from present experimental data.
NLO corrections are always pointing in the right direction of closer
agreement with experiment.  In particular, the energy dependence of the gluon
jet average multiplicity, of the ratio of its slopes for gluon and quark jets
and the values of widths at $Z^0 $-energies can be fitted with rather high
accuracy. However, it still fails in the ratio of average multiplicities of
gluon and quark jets differing from experiment by about 30$\%$ in absolute
values and even more in its energy dependence slope \cite{cd}. Here,  2NLO
and 3NLO terms improve the situation so that the ratio $r$ differs by about
10--15$\%$ only. They do not spoil good qualitative features of NLO in the
energy dependence \cite{dg}.
However, these corrections become very large for the widths and for the third
moments as shown above, as well as for the slope and curvature in the energy
dependence of average multiplicities \cite{cd}.

Thus we conclude that present perturbative QCD results can describe the
experimental data within 10--15$\%$ accuracy. The perturbative series breaks
down, however, at different orders for different quantities.
There seems to be no standard way to improve the analytic results and
unanimously predict where one should truncate the expansion.
Nevertheless, the general trends obtained from the
perturbative approach are steadily indicating qualitative convergence of
theory and experiment. Moreover, the computer results \cite{lo} show that
the exact solutions of QCD equations can be even closer to experiment.

\acknowledgements
This research is supported in part by the Natural Sciences and
Engineering Research Council of Canada, and the Fonds pour la Formation de
Chercheurs et l'Aide \`a la Recherche of Qu\'ebec, and also by INTAS
and Russian Foundation for Basic Research.

\appendix
\section*{}
The terms proportional to $\langle n_G\rangle ^2z^2/2$ in the right-hand sides
of equations (\ref{gg''}), (\ref{gf''})  up to $O(\gamma _0^3)$  corrections
can be written as
\begin{eqnarray}
G_G-1\rightarrow F_2^G=\frac {4}{3}(1-f_1\gamma _0-f_2\gamma_0^2-f_3\gamma_0^3),
\\
G_G^{'}\rightarrow 2\gamma F_2^G+F_2^{G'}=\frac {8\gamma_0}{3}[1-(a_1+f_1)
\gamma_0+
\nonumber
\\
\gamma_0^2(a_1f_1-a_2-f_2+0.5Bf_1)],
\\
G_G^{''}\rightarrow 2(2\gamma^2+\gamma^{'})F_2^G+4\gamma F_2^{G'} +F_2^{G''}=
\frac{16}{3}\gamma_0^2[1-\gamma_0(2a_1+f_1+0.5B)],
\\
G_G^{'''}\rightarrow 8\gamma^3F_2^G=\frac {32}{3}\gamma_0^3,
\\
\left (\frac {G_G^{'2}}{G_G}\right )'\rightarrow 4\gamma_0^3,
\\
\left (\frac {G_F^2}{G_G}\right )'\rightarrow \gamma_0 \frac{4}{3r_0^2}
(r_0^2-5r_0+6)+\gamma_0^2[\frac {8(f_1+a_1)}{3}-
\nonumber
\\
4\frac {1+r_0/3}{r_0^2}(\phi_1+a_1-2r_1)-4a_1 (1-r_0^{-1})^2
-\frac {8r_1}{r_0}(1-r_0^{-1})]+
\nonumber
\\
2\gamma_0^3r_0^{-2}[2(r_0-1)(2a_1r_1+Br_1-2(r_2+r_1^2))-2a_2(r_0-1)^2+2r_1^2-
\nonumber
\\
\frac {4r_0^2}{3}(a_1f_1-a_2-f_2)+B((1+r_0/3)(\phi_1-2r_1)-\frac {2}{3}
f_1r_0^2)+
\nonumber
\\
2(1+r_0/3)(2r_2-a_2-\phi_2+3r_1^2+a_1\phi_1-2a_1r_1-2\phi_1r_1)],
\\
\left ( \frac {G_FG_F'}{G_G}\right )'\rightarrow  8\gamma_0^2r_0^{-2}
(1-r_0/3)+4\gamma_0^3
r_0^{-2}[4r_1-\frac {r_0r_1}{3}-
\nonumber
\\
(1+r_0/3)\phi_1-(1-r_0/3)(4a_1+B)],
\\
\left ( \frac {G_FG_F^{''}}{G_G}\right )'\rightarrow \gamma_0^3\frac
{4(9-r_0)}{3r_0^2},
\\
\left (\frac {G_F^{'2}}{G_G}\right )'\rightarrow \gamma_0^3\frac {4}{r_0^2},
\\
\left ( \frac {G_GG_F'}{G_F}\right )'\rightarrow \gamma_0^2\frac {16}{3r_0^2}
+\gamma_0^3
\frac{4}{3r_0}[5r_1-8a_1-2B-(1+3r_0^{-1})\phi_1],
\\
\left (\frac {G_G'G_F'}{G_F}\right )'\rightarrow \gamma_0^3\frac {4}{r_0},
\\
\left (\frac {G_GG_F^{''}}{G_F}\right )'\rightarrow \gamma_0^3 4r_0^{-2}
(1+5r_0/3),
\\
\frac {G_F^2}{G_G}-1\rightarrow \frac {2}{3r_0^2}(r_0^2-5r_0+6)+\gamma_0
\frac {2}{3r_0^2}
(2f_1r_0^2-(3+r_0)\phi_1+4r_1(3-r_0)),
\\
\frac {G_FG_F'}{G_G}\rightarrow \gamma_04r_0^{-2}(1-r_0/3),
\\
\frac {G_GG_F'}{G_F}\rightarrow \gamma_0\frac {8}{3r_0}.
\end{eqnarray}

The coefficients $d_i$ in (\ref{f3}), (\ref{p3}) are
\begin{equation}
d_1=2(2a_1f_1-2a_2-2f_2+3Bf_1),
\end{equation}
\begin{equation}
d_2=2(4a_1+2f_1+5B),
\end{equation}
\begin{equation}
d_3=-4,
\end{equation}
\begin{eqnarray}
d_4=-\frac {n_f}{2N_cr_{0}^{2}}[r_0^2f_2-0.5(3+r_0)\phi _2-r_0^2f_1(a_1+1.5B)+
\nonumber
\\
0.25(3+r_0)\phi _1(2a_1+3B-4r_1)+r_0^2a_2-0.5(3+r_0)a_2+2r_2(3-r_0)+
\nonumber
\\
r_1[r_0(2a_1-1.5r_1+3B)-6a_1+9r_1-9B]-1.5a_2(r_0-1)^2],
\end{eqnarray}
\begin{equation}
d_5=\frac {n_f}{2N_cr_0^2}[(3+r_0)\phi _1+4(3-r_0)a_1+B(21-13r_0+2r_0^2)-
r_1(12-r_0)],
\end{equation}
\begin{equation}
d_6=-\frac {n_f(9-r_0)}{4N_cr_0^2},
\end{equation}
\begin{equation}
d_7=0.5d_1,
\end{equation}
\begin{equation}
d_8=-d_2,
\end{equation}
\begin{equation}
d_9=-d_3,
\end{equation}
\begin{equation}
d_{10}=-r_0^{-1}[8a_1+(1+3r_0^{-1})\phi _1+2B(3+2r_0)-5r_1],
\end{equation}
\begin{equation}
d_{11}=\frac {3+5r_0}{2r_0^2},
\end{equation}
\begin{equation}
d_{12}=-3,
\end{equation}
\begin{equation}
d_{13}=-\frac {3n_f}{4N_cr_0^2},
\end{equation}
\begin{equation}
d_{14}=3/r_0.
\end{equation}

The integrals $\vv_i$ are as follows
\begin{eqnarray}
\vv_1=\int _0^1Vdx=\int _0^1(1-\frac {3}{2}x+x^2-\frac {x^3}{2})dx=
\frac {11}{24}, \\
\vv_2=\int_0^1 \left [\frac {\ln (1-x)}{x}-2V\ln x(1-x)\right ]dx=
\frac {67-6\pi ^{2}}
{36},  \\
\vv_3=\int_0^1 \left [\frac {\ln ^{2}(1-x)}{x}-2V(\ln ^{2}x+\ln ^{2}(1-x))
\right ]dx=
2\zeta (3)-\frac {413}{108},   \\
\vv_4=\int_0^1 [x^2+(1-x)^2]dx=\frac {2}{3} , \\
\vv_5=\int _0^1[x^2+(1-x)^2]\ln x dx=-\frac {13}{18}, \\
\vv_6=\int_0^1 [x^2+(1-x)^2]\ln ^{2}x dx=\frac {89}{54},  \\
\vv_7=\int _0^1\Phi dx=\int (1-\frac {x}{2})dx=\frac {3}{4},  \\
\vv_8=\int _0^1\Phi \ln x dx=-\frac {7}{8},  \\
\vv_9=\int _0^1\Phi \ln ^{2}x dx= \frac {15}{8},  \\
\vv_{10}=\int_0^1 \left [\Phi -\frac {1}{x}\right ]\ln (1-x) dx=\frac
{\pi ^2}{6}-
\frac {5}{8},  \\
\vv_{11}=\int_0^1 \left [\Phi-\frac {1}{x}\right ]\ln ^{2}(1-x) dx =\frac {9}{8}
-2\zeta (3),\\
\vv_{12}=\int _0^1 (x^{-1}-2V)\ln x\ln (1-x) dx=\zeta (3)-395/216+11
\pi ^2/72 ,\\
\vv_{13}=\int _0^1 [x^2+(1-x)^2]\ln x\ln (1-x) dx=71/54-\pi ^2/9 ,\\
\vv_{14}=\int _0^1 (1-0.5x-x^{-1})\ln x\ln (1-x) dx=1.5-\pi ^2/8-\zeta (3).
\end{eqnarray}
$\zeta  $ means Riemann's $\zeta  $-function.

\begin{figure}[h]
\vspace*{12cm}
\includegraphics{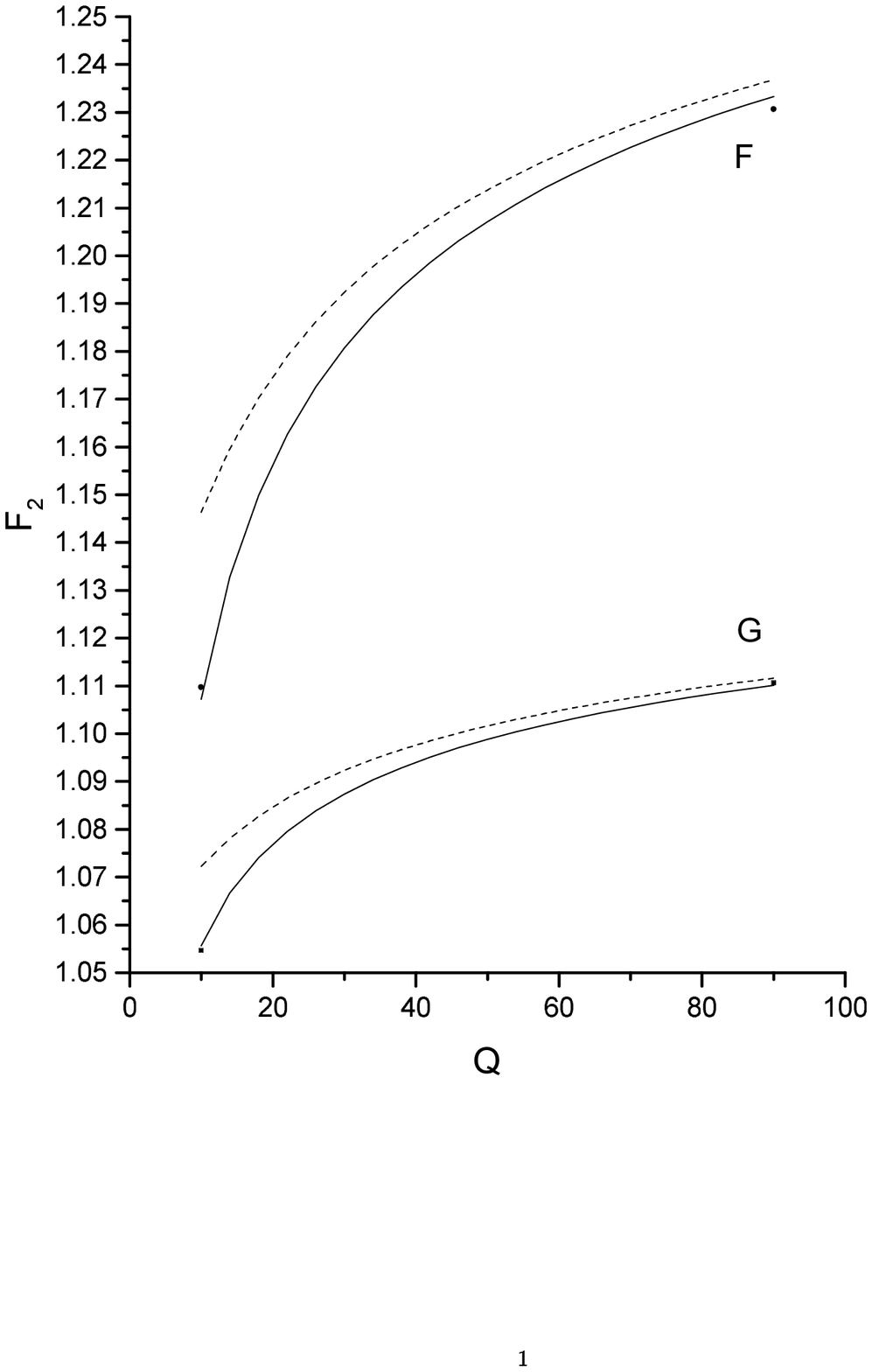}
\vspace*{5cm}
\caption{The energy behaviour of the second factorial moments of quark (F) and
gluon (G) jets. The limits of integration are chosen as 0 and 1 (solid lines)
or $\varepsilon $ and $1-\varepsilon $ (dashed lines). The dots at the ends
show that the curves are insensitive to variation of the effective number of
flavors (see text).}{}
\end{figure}

\end{document}